\input amstex
\input xy
\xyoption{all}
\documentstyle{amsppt}
\document
\magnification=1200
\NoBlackBoxes
\nologo
\hoffset1.5cm
\voffset2cm

\pageheight {16cm}

\bigskip

\centerline{\bf A COMPUTABILITY CHALLENGE: ASYMPTOTIC BOUNDS}
\medskip
\centerline{\bf AND ISOLATED ERROR--CORRECTING CODES}

\bigskip

\centerline{\bf Yuri I.~Manin}

\medskip

\centerline{\ Max--Planck--Institut f\"ur Mathematik, Bonn, Germany}
\bigskip

\hfill{\it Dedicated to Professor C.~S.~Calude, on his 60th birthday}
\bigskip

ABSTRACT. Consider the set of all error--correcting block codes over a fixed
alphabet  with $q$ letters. It determines a recursively enumerable set of points
in the unit square with coordinates $(R,\delta )$:= {\it (relative transmission rate,
relative minimal distance).} Limit points of this set form a closed subset,
defined by $R\le \alpha_q(\delta )$, where $\alpha_q(\delta )$
is a continuous decreasing function called {\it asymptotic bound.}
Its existence was proved by the author in 1981, but all attempts to
find an explicit formula for it so far failed.

In this note I consider the question whether this function is computable
in the sense of constructive mathematics, and discuss some arguments
suggesting that the answer might be negative.

\bigskip
\centerline{\bf 1. Introduction.}

\medskip

{\bf 1.1. Notation.} This paper is a short survey focusing  on an unsolved problem of
the theory of error--correcting codes (cf. the monograph  [VlaNoTsfa]).
\smallskip

Briefly, we choose and fix an integer $q\ge 2$ and a finite set, {\it alphabet} $A$,
of cardinality $q$. An (unstructured) {\it code} $C$ is defined as a non--empty subset $C\subset A^n$
of words of length $n\ge 1$.  Such $C$ 
determines its {\it code point} $P_C= (R(C),\delta (C))$ in the $(R,\delta )$--plane,  where $R(C)$ is
called  {\it the transmission rate}
and $\delta (C)$ is {\it the relative minimal distance of the code.}  They are defined by
the formulas
$$
\delta (C):= \frac{d(C)}{n(C)}, \quad  
d(C) := \roman{min}\,\{d(a,b)\,|\,a,b\in C, a\ne b\},\quad n(C):=n,
$$
$$
R(C) = \frac{k(C)}{n(C)}, \quad  k(C):=\roman{log}_q \roman{card}(C),
\eqno(1.1)
$$

where $d(a,b)$ is the Hamming distance
$$
d((a_i),(b_i)):= \roman{card} \{i\in (1,\dots ,n)\,|\,a_i\ne b_i\}.
$$
In the degenerate case $\roman{card}\,C=1$ we put $d(C)=0.$
We will call the numbers  $k=k(C)$, $n=n(C)$, $d=d(C)$, {\it code parameters}
and refer to $C$ as an $[n,k,d]_q$--code. 
\smallskip

A considerable bulk of research in this domain
is dedicated either to the construction of (families of) ``good'' codes (e.~g. algebraic--geometric ones),
or to the proof that ``too good'' codes do not exist.  A code is good if   in a sense it maximizes 
simultaneously the transmission rate and the minimal distance. To be useful in applications, 
a good code must
also come with feasible algorithms of encoding and decoding. The latter task includes
the problem of finding a closest (in Hamming's metric) word in $C$, given an
arbitrary word in $A^n$ that can be an output of a noisy transmission channel (error correction). Feasible algorithms exist for certain classes
of {\it structured} codes. The simplest and most popular example is that
of {\it linear codes}: $A$ is endowed with a structure of a finite field $\bold{F}_q$,
$A^n$ becomes a linear space over $\bold{F}_q$, and $C$ is required to be a linear subspace.

\medskip

{\bf 1.2.  Asymptotic bounds.} Since the demands of good codes are mutually conflicting, it is natural to look
for the bounds of possible.

\smallskip
A precise formulation
of the notion of good codes can be given in terms of two notions:
{\it asymptotic bounds} and {\it isolated codes.} 

\smallskip
Fix $q$ and  denote by $V_q$ the set of all points $P_C$, corresponding to all
$[n,k,d]_q$--codes. Define {\it the code domain} $U_q$ as {\it the set of limit points of $V_q$.}

It was proved in [Man1] that   $U_q$ consists of all points in $[0,1]^2$ lying below the graph of a certain continuous
decreasing function  $\alpha_q$:
$$
U_q=\{(R,\delta )\,|\,R\le \alpha_q(\delta )\}.
\eqno(1.2)
$$
Moreover, $\alpha_q(0)=1, \alpha_q(\delta )=0$ for $1-q^{-1}\le \delta \le 1$,
and the graph of $\alpha_q$ is tangent to the $R$--axis at $(1,0)$ and to the $\delta$--axis
at $(0,1-q^{-1})$.
\smallskip
This curve is called {\it the asymptotic bound}. (In fact, [Man1] considered only linear codes,
and the respective objects are now called $V_q^{lin}, U_q^{lin}, \alpha_q^{lin}$; unstructured case can be treated in the same way with minimal changes:
cf. [ManVla] and [ManMar]).

\smallskip

Now, a code can be considered a good one, if its point either lies in $U_q$ and is close
to the asymptotic bound, or is {\it isolated}, that is, lies above the asymptotic bound.

\medskip

{\bf 1.3. Computability problems.} There is an abundant literature 
establishing upper and lower estimates for
asymptotic bounds, and providing many isolated codes.
However, not only ``exact formulas'' for asymptotic bounds are unknown, but even
the question, whether $\alpha_q(\delta )$ is differentiable, remains open
(of course, since this function is monotone and continuous, it is differentiable {\it almost everywhere}.)
Similarly, the structure of the set of  isolated code points  
 is a mystery: for example, {\it are there points on $R=\alpha_q(\delta ),
0<R<1-q^{-1}$, that are limit points of isolated codes?}

\smallskip

The principle goal of this report is to discuss weaker versions of these problems, 
replacing  ``exact formulas''  by ``computability''.  In particular, we try to elucidate
the following 

\medskip

{\it QUESTION. Is the function $\alpha_q(\delta )$ computable?}
\medskip

As our basic model of computability we adopt the one described in [BratWe] and further developed
in [BratPre], [Brat],  [BratMiNi].  In its simplest concrete version, it involves
approximations of closed subsets of $\bold{R}^2$, such as $U_q$ or graph of $\alpha_q$,
by  unions of computable sets of rational coordinate squares, ``pixels''
of varying size.

\smallskip

The following mental experiment suggests that the answer to this computability problem
may not be obvious, and that  $\alpha_q$ might  even be {\it uncomputable} and by implication
not expressible by any reasonable  ``explicit formula''.  

\smallskip

Imagine that a computer is drawing
finite approximations $V_q^{(N)}$ to the set of code points $V_q$ by plotting all points
with $n\le N$ for a  large $N$ (appropriately matching a chosen pixel size). What will we see on the screen? 

\smallskip

Conjecturally, we will {\it not} see  a dark domain approximating $U_q$ with a cloud of isolated
points above it, but rather an eroded version of the {\it Varshamov--Gilbert curve} lying
(at least partially) strictly below $R=\alpha_q(\delta )$:
$$
R= \frac{1}{2}(1-\delta\roman{log}_q (q-1)   - \delta\roman{log}_q \delta -
(1-\delta )\roman{log}_q (1-\delta ) )
\eqno(1.3)
$$
In fact, ``most''  code points lie ``near'' (1.3): cf. Exercise 1.3.23 in [VlaNoTsfa]
 and some precise statements in [BaFo] (for $q=2$.)

\smallskip

By contrast, a statistical meaning of the asymptotic bound does not seem to be known,
and this appears as the intrinsic difficulty  for a complete realization of the project
started in [ManMar]: interpreting asymptotic bound as a ``phase transition'' curve.
Hopefully, a solution might be found if we imagine plotting
code points in the order of their {\it growing Kolmogorov complexity},
as was suggested and used in [Man3] for renormalization of halting problem.
For the context of constructive mathematics,  cf. [CaHeWa] and references therein.

\smallskip

In any case, it is clear that code domains represent an interesting
testing ground for various versions of  computability of subsets of $\bold{R}^n$,
complementing the more popular Julia and Mandelbrot fractal sets (cf. [BravC] and [BravYa]).

\bigskip

\centerline{\bf 2. Code parameters and code points: a summary}

\medskip

{\bf 2.1. Constructive worlds of code parameters.} Denote the set of all triples
$[n, q^k, d]\in \bold{N}^3$ corresponding to all (resp. linear) $[n,k,d]_q$--codes
by $P_q$ (resp. $P^{lin}_q$). Clearly, $P_q$ and $P^{lin}_q$ are infinite decidable subsets of  $\bold{N}^3$.
Therefore they admit natural recursive and recursively invertible bijections with $\bold{N}$
(``admissible numberings''),
defined up to composition with any  recursive
permutation $\bold{N}\to \bold{N}$. Hence   $P_q$ and  $P^{lin}_q$ are  infinite
{\it constructive worlds} in the sense of [Man3], Definition 1.2.1.
\smallskip

If $X$, $Y$ are two constructive worlds, we can unambiguously define the notions of
(partial) recursive maps $X\to Y$, enumerable and decidable subsets of $X$, $Y$,
$X\times Y$ etc., simply pulling them back to the numberings. For a more developed
categorical formalism, cf. [Man3].

\medskip

{\bf 2.2. Constructive world $\bold{S}=[0,1]^2\cap \bold{Q}^2$.} The set of all rational
points of the unit square in the $(R,\delta )$--plane also has a canonical structure of a 
constructive world. 

\medskip

{\bf 2.3.  Enumerable sets of code points.}  Code points (1.1) of linear codes all lie in $\bold{S}$.
To achieve this for unstructured codes, we will slightly amend (1.1) and define
the map $cp:\ P_q\to \bold{S}$ ($cp$ stands for ``code point'') by
$$
cp([n,q^k,d]):= \left(\frac{[k]}{n}, \frac{d}{n}\right)
\eqno(2.1)
$$
where $[k]$ denotes the integer part of the (generally real) number $k$.
On $P_q^{lin}\subset P_q$ it coincides with (1.1). 

\smallskip
The motivation for choosing (2.1) is this: in the eventual study of computability properties
of the graph $R=\alpha_q(\delta )$, it is more transparent to approximate
it by points with rational coordinates, rather than logarithms.

\smallskip

Let $V_q$ (resp. $V_q^{lin}$) be the image $cp(P_q)$ (resp. $cp(P_q^{lin}$)) i.e.  the 
respective set of code points
in $\bold{S}$. Since $cp$ is a total recursive function both on $P_q$ and  $P_q^{lin}$,
$V_q$ and $V_q^{lin}$ are recursively enumerable subsets of  $\bold{S}$.

\medskip

{\bf 2.4. Limit code points.} Let $U_q$ (resp. $U_q^{lin}$) be the closed sets of limit
points of $V_q$ (resp.  $V_q^{lin}$).  We will call {\it limit code points}  elements of $V_q\cap U_q$
(resp.  $V_q^{lin}\cap U_q^{lin}$). The remaining subset of {\it isolated code points} is defined as $V_q\setminus V_q\cap U_q$, and similarly for linear codes.
\smallskip

Notice that  we get one and the same set $U_q$, using transmission rates (1.1) or (2.1).
In fact, for any infinite sequence of pairwise distinct code parameters $[n_i,q^{k_i},d_i]$,
$i=1,2, ...$ we have $n_i\to \infty$, hence the convergence of the sequence
of code points (1.1) is equivalent to that of (2.1), and they have a common limit.
The resulting sets of isolated code points
differ depending on the adopted definition (1.1) or (2.1), however, the set of
{\it isolated codes}, those whose code points are isolated, remains the same.

\smallskip

Our main result in this section is the following characterization of
limit and isolated code points
in terms of the recursive map $cp$ rather
than topology of the unit square.

\smallskip

We will say that a code point $x\in V_q$ has {\it infinite} (resp. {\it finite}) {\it multiplicity},
if $cp^{-1}(x)\subset P_q$ is infinite (resp. finite).
The same definition applies to $V_q^{lin}$ and $P_q^{lin}$.

\medskip

{\bf 2.5. Theorem.} {\it (a) Code points of infinite multiplicity are limit points.
Therefore isolated code points have finite multiplicity.

\smallskip

(b) Conversely, any  point $(R_0,\delta_0)$ with rational coordinates
satisfying the inequality $0<R_0<\alpha_q(\delta_0)$ 
(resp. $0<R_0<\alpha_q^{lin} (\delta_0)$) is a code point (resp. linear code point) of infinite multiplicity.}

\medskip

This (actually, a slightly weaker) statement, seemingly, was first stated  and proved in [ManMar]. It makes me suspect
that {\it distinguishing between limit and isolated code points might be
algorithmically undecidable}, since in general it is algorithmically
impossible to decide, whether  a given recursive function takes
one of its values at a finite or infinitely many points.

\smallskip

Similarly, one cannot expect {\it a priori} that limit and isolated code points
form two recursively enumerable sets, but this must be true,
if $\alpha_q$ is computable: see Theorem 3.3.1 below.

\smallskip

For completeness, I will reproduce the proof  of Theorem 2.5 here. It is
based on the same ``Spoiling Lemma'' that underlies the only known 
proof of existence of the asymptotic bounds
$\alpha_q$ and $\alpha_q^{lin}.$

\medskip
{\bf 2.6. Proposition} (Numerical spoiling). {\it  If there exists
a linear $[n,k,d]_q$--code, then there exist also linear
codes with the following parameters:

\smallskip

(i) $[n+1,k,d]_q$  (always).

\smallskip

(ii) $[n-1,k,d-1]_q$ (if  $n>1,k>0$.)

\smallskip

(iii) $[n-1,   {k-1}, d]_q$ (if  $n>1$, $k>1)$  
\smallskip

In the domain of unstructured codes statements (i) and (ii) remain true, whereas
in (iii) one should replace $[n-1,   {k-1}, d]_q$  by  $[n-1,   {k^{\prime}}, d]_q$  for some $k-1\le k^{\prime} <k$.}

\medskip

For a proof of Proposition 2.6, see e.~g. [VlaNoTsfa] (linear codes)
and [ManMar] (unstructured codes).

\medskip

{\bf 2.7. Proof of Theorem 2.5.} (a)  We first check that if a code point  $(R_0,\delta_0 )\in \bold{Q}^2$
is of infinite multiplicity, then it is a limit point.
In fact, let $[n_i, q^{k_i},d_i]$ be an infinite sequence of pairwise distinct code parameters, $i\ge 1$,
such that $[k_i]/n_i=R_0, d_i/n_i=\delta_0$ for all $i$. Then codes with parameters
$[n_i+1, q^{k_i},d_i]$ (cf. 2.6 (i)) produce infinitely many pairwise distinct code points
converging to $(R_0,\delta_0 )$.

\smallskip

(b) Now  consider a  rational point $(R_0,\delta_0 )\in \bold{Q}^2\cap (0,1)^2$
(unstructured or linear), lying strictly below the respective asymptotic bound.
Then there exists a code point $(R_1,\delta_1)$ also lying strictly
below the asymptotic bound, with $R_1>R_0$ and $\delta_1>\delta_0$, because 
functions $\alpha_q$ and $\alpha_q^{lin}$ decrease. Hence in the
part of $U_q$ (resp. $U_q^{lin}$) where $R\ge R_1,\delta\ge \delta_1)$
there exists an infinite family of pairwise distinct code points $(R_i,\delta_i)$, $i\ge 1$,
coming from a family of unstructured (resp. linear) $[N_i, {K_i}, D_i]_q$--codes.

\smallskip

Let  $(R_0 ,\delta_0)= (k/n, d/n)$. Divide $N_i$ by $n$ with a remainder term, i.e.
put $N_i=(a_i-1)n+r_i$, $a_i\ge1, 0\le r_i <n.$ Using repeatedly 2.6 (i), spoil the respective
$[N_i, {K_i}, D_i]_q$--code, replacing it by some  $[a_in, {K_i}, D_i]_q$--code.
Its code point will have slightly smaller coordinates than the initial $(R_i,\delta_i)$,
however for $N_i$ large enough, it will remain in the domain $R>R_0,\delta >\delta_0$.
Hence we may and will assume from the start that in our sequence
of  $[N_i, {K_i}, D_i]_q$--codes all $N_i$'s are divisible by $n$: 
$$
N_i=a_in \, .
\eqno(2.2)
$$
In order to derive  by spoiling from this sequence another sequence of pairwise distinct codes,
all of which have one and the same code point $(R_0,\delta_0)=(k/n,d/n)$, we will first
consider the case of linear codes where the procedure is neater, because $[K_i]=K_i$.
Since we have $K_i/N_i>k/n, D_i/N_i>d/n$, we get
$$
K_i>a_ik,\quad  D_i>a_id.
$$
To complete the proof, it remains to reduce the parameters $K_i,D_i$
to $a_ik, a_id$ respectively, without reducing $N_i=a_in.$ 
In the linear case, this is achieved by application of several steps 2.6 (ii), 2.6 (iii),
followed by steps 2.6 (i).

\smallskip

In the unstructured case reducing $D_i$ can be done in the same way.
It remains to reduce $[K_i]$ to $a_ik$.
 One application of the step
2.6 (iii) produces $K_i^{\prime}$ such that either $[K_i^{\prime}]= [K_i]-1$,
or  $[K_i^{\prime}]= [K_i]$. In the latter case, after restoring $N_i$ to its former value, one must
apply 2.6 (iii) again. After a finite number of such substeps, we will finally get $[K_i]-1$.

\medskip

{\bf 2.8. Question.} {\it Can one find a recursive function
$b(n,k,d,q)$ such that if an $[n,k,d]_q$--code is isolated,
and $a>b(n,k,d,q)$, there is no code with parameters
$[an,ak,ad]_q$?}

\bigskip

\centerline{\bf 3. Codes and computability}

\medskip

In this section, I will discuss  computability of two types of closed sets
in $[0,1]^2$: $U_q$ and $\Gamma_q$:= the graph of $\alpha_q$,
as well as their versions for linear codes. I will start with the brief
summary of basic definitions of [BratWe] in our context.

\medskip

{\bf 3.1. Effective closed sets.} First, we will consider $[0,1]^2$, $U_q$ and $\Gamma_q$ as  closed subsets
in a larger square, say $X:=[-1,2]^2$, with its structure of compact metric space
given by $d((a_i),(b_i)):=\roman{max}\, |a_i-b_i|$.  The set of {\it open balls $\Cal{B}$ with
rational centers and radii} in this space has a natural structure of a constructive world
(cf. 2.1). Hence we may speak about (recursively) enumerable and decidable subsets
of $\Cal{B}$.

\smallskip

Following [BratWe] and [La], we will consider three types of effectivity of closed subsets  $Y\subset X$:

\smallskip

(i) $Y$ is called {\it recursively enumerable}, if the subset
$$
\{I\in \Cal{B}\,|\, I\cap Y\ne \emptyset\}\subset \Cal{B}
\eqno(3.1)
$$
is recursively enumerable in $\Cal{B}$.

\medskip

(ii) $Y$ is called {\it co--recursively enumerable}, if the subset
$$
\{I\in \Cal{B}\,|\, \overline{I}\cap Y= \emptyset\}\subset \Cal{B}
\eqno(3.2)
$$
is recursively enumerable in $\Cal{B}$ (here $\overline{I}$ is the closure of $I$).
\medskip

(iii) $Y$ is called {\it recursive}, if it is simultaneously  recursively enumerable
and  co--recursively enumerable.

\medskip

As a direct application of [BratWe] we find:

\medskip

{\bf  3.2. Proposition.} {\it The closures  $\overline{V}_q$ and $\overline{V}_q^{lin}$ are 
recursively enumerable.}

\smallskip

{\bf Proof.} In fact, range of the function $cp$ (see 2.3) is dense in
 $\overline{V}_q$, resp.  $\overline{V}_q^{lin}$, and we can apply  [BratWe],
 Corollary 3.13(1)(d).

\medskip

{\bf 3.3. Problem of computability of the asymptotic bound.} Referring to the Corollary 7.3 of [Brat], we
will call $\alpha_q$ (resp. $\alpha_q^{lin}$) {\it computable}, if its graph
$\Gamma_q$ (resp. $\Gamma_q^{lin}$) is co--recursively enumerable.

\medskip

{\bf 3.3.1. Theorem.} {\it Assume that $\alpha_q$ is computable.  Then
each of the following sets is recursively enumerable:

\smallskip

(a) Code points lying strictly below the asymptotic bound.

\smallskip

(b) Isolated code points.

\smallskip

The same is true for linear codes, if $\alpha_q^{lin}$ is computable.}

\medskip

{\bf Proof.}  We start with the following remark. Choose any integer $N\ge1$
and consider the set $\Gamma_q^{(N)}$ which is the union 
of closed balls of the form 
$$
\overline{I}= \left[ \frac{p}{N},\frac{p+1}{N}\right]\times \left[\frac{p}{N},\frac{p+1}{N}\right] \subset X
\eqno(3.3)
$$
satisfying $p\in \bold{N}$, $\overline{I}\cap \Gamma_q\ne \emptyset$. Then we have:
\smallskip
(i) {\it The boundary of
$\Gamma_q^{(N)}$ consists
of two vertical (parallel to the $R$--axis)  segments at the ends and two piecewise linear connected closed curves:
$\Gamma_{q+}^{(N)}$ lying above   $\Gamma_{q-}^{(N)}$.

\smallskip

(ii) The distance of any point $x\in\Gamma_{q-}^{(N)}$ 
to  $\Gamma_{q+}^{(N)}$ does not exceed $2/N$, and similarly with $+$ and $-$ reversed.}

\smallskip

Let us call an $N$--{\it strip} any connected closed set satisfying these conditions.

\smallskip

Now, assuming $\alpha_q$ (resp. $\alpha_q^{lin}$) computable,
that is, $\Gamma_q$ co--recursively enumerable,  choose $N$ and
run the algorithm generating
in some order all rational closed balls $\overline{I}$ such that  $\overline{I}\cap \Gamma_q=\emptyset$.
Wait until their subset consisting of balls of the form (3.3) covers the whole square $[0,1]^2$ with exception of a set whose closure is an $N$--strip.
This strip will then be an approximation to
$\Gamma_q$ (resp. $\Gamma_q^{lin}$) containing the respective graph in the subset of its
inner points.

\smallskip

Run parallelly  an algorithm generating all code points and divide each partial list of code points
into three parts depending on $N$: points lying below  $\Gamma_q^{(N)}$,
above  $\Gamma_q^{(N)}$, and inside $\Gamma_q^{(N)}$.

\smallskip

When $N$ grows, the growing first and second parts respectively will recursively enumerate code points
below and above the asymptotic bound.

\smallskip

{\bf Remark.} This reasoning also shows, in accordance with [Brat],
that if we assume $\Gamma_q$ only co--recursively enumerable, it will
be automatically recursively enumerable and therefore recursive.

\medskip

{\bf 3.4. Theorem.}  {\it Assume that $U_q$ is recursive in the sense of 3.1(iii).
Then $\alpha_q$ is computable. The similar statement holds for linear codes.}

\medskip

{\bf Proof.} Consider first a closed ball $\overline{I}$  as in (3.3) 
that intersects $U_q$ whereas its inner part $I$ does
not intersect $U_q$. A contemplation will convince the reader that
the left lower boundary point of this ``ball'' (a square in the Euclidean metric)
is precisely the intersection point $\overline{I}\cap \Gamma_q$.
Call such a ball {\it an exceptional $N$--ball.} Since $\alpha_q$ is decreasing, we have
\medskip

(a) {\it Each horizontal strip $p/N\le R\le (p+1)/N$ and each vertical
strip   $q/N\le \delta \le (q+1)/N$ can contain no more than one  exceptional $N$--ball.

\medskip

(b) If one exceptional $N$--ball lies to the right of another one, then it also lies lower than that one.}

\medskip

Generally, call a set of $N$--balls {\it $N$--admissible}, if it satisfies (a) and (b).

\smallskip
Now, assuming $U_q$ recursive and having chosen $N$, we can  run parallelly two algorithms:
one generating closed balls (3.3) non--intersecting $U_q$ and another,
generating open balls (3.3) intersecting $U_q$. Run them until
all $N$--balls are generated, with a possible exception of
an $N$--admissible subset $X_q^{(N)}$, then stop generation. Let $U_{q+}^{(N)}$ be the union
of all balls generated by the first algorithm, and $U_{q-} ^{(N)}$  the union
of all balls generated by the second algorithm.

\smallskip

Look through all the balls in $X_q^{(N)}$ in turn. If there are elements in it
whose closure does not intersect the closure of   $U_{q-}^{(N)}$,
delete them from  $X_q^{(N)}$ and put it into  $U_{q+}^{(N)}$. Similarly,
if there are elements in it
whose closure does not intersect  (initial)  $U_{q+}^{(N)}$,
delete them from  $X_q^{(N)}$ and put them into  $U_{q-}^{(N)}$.
 
\smallskip

Keep the old notations   $U_{q-}^{(N)}$,
 $U_{q+}^{(N)}$,  $X_q^{(N)}$ for these amended sets.

\smallskip
Now,  the union of the lower boundary of  $U_{q+}^{(N)}$ and the upper boundary of   $U_{q-}^{(N)}$
will approximate $\Gamma_q$ from two sides, with error not exceeding
$N^{-1}$.  (Here a "boundary" means the respective set of  boundary squares).
\smallskip

Clearly, this reasoning shows also also computability of $\alpha_q$ in the sense of  3.3.

\bigskip
\centerline{\bf References}

\medskip

[BaFo] A.~Barg, G.~D.~Forney. {\it Random codes: minimum distances and error exponents.}
IEEE Transactions on Information Theory, vol 48, No 9 (2002), 2568--2573.

\smallskip

[Brat] V.~Brattka.  {\it Plottable real functions and the computable graph theorem.}
SIAM J. Comput., vol. 38, Bo. 1 (2008), 303--328.

\smallskip

[BratMiNi] V.~Brattka, J.~S.~Miller, A.~Nies. {\it Randomness and differentiability.}
arXiv:1104.4456
\smallskip

[BratPre] V.~Brattka, G.~Preser. {\it Computability on subsets of metric spaces.}
Theoretical Computer Science,
305 (2003), 43--76.
\smallskip

[BratWe] V.~Brattka, K.~Weihraub. {\it Computability on subsets of Euclidean
space I: closed and compact subsets.} Theoretical Computer Science,
219 (1999), 65--93.

[BravC]  M.~Braverman, St.~Cook. {\it Computing over the reals: foundations for scientific
computing.} Notices AMS, 53:3 (2006), 318--329

\smallskip

[BravYa] M.~Braverman, M.~Yampolsky. {\it Computability of Julia sets.} Moscow Math. Journ., 8:2 (2008), 185--231.

\smallskip

[CaHeKhWa]   C.~S.~Calude, P.~Hertling, B.~Khoussainov, Yongge Wang.
{\it Recursively enumerable reals and Chaitin $\Omega$ numbers.}
Theor. Comp. Sci, 255 (2001), 125--149.

\smallskip

[La]  D.~Lacombe. {\it Extension de la notion de fonction r\'ecursive aux fonctions d'une ou plusieurs
variables r\'eelles., I--III.}\  C.~R.~Ac.~Sci.~Paris, 240 (1955), 2478--2480;
241 (1955), 13--14, 151--153.

\smallskip

[Man1] Yu.~I.~Manin, {\it What is the maximum number of points
on a curve over $\bold{F}_2$?} J. Fac. Sci. Tokyo, IA, Vol. 28 (1981),
715--720. 

\smallskip

[Man2] Yu.~I.~Manin, {\it  Renormalization and computation I: motivation and background.}
Preprint math.QA/0904.4921
\smallskip
[Man3] Yu.~I.~Manin, {\it  Renormalization and Computation II: Time Cut-off and the Halting Problem.}
Preprint math.QA/0908.3430

\smallskip

[ManMar]  Yu.~I.~Manin, M.~Marcolli. {\it Error--correcting codes and phase transitions.}
arXiv:0910.5135

\smallskip

[ManVla] Yu.~I.~Manin. S.G.~Vladut, {\it Linear codes and
modular curves}. J. Soviet Math., Vol. 30 (1985), 2611--2643.
\smallskip

[TsfaVla] M.~A.~Tsfasman, S.~G.~Vladut. {\it Algebraic--geometric
codes}, Kluwer, 1991.

\smallskip

[VlaNoTsfa] S.~G.~Vladut, D.~Yu.M.~A.~Tsfasman. {\it  Algebraic geometric codes: basic notions.} Mathematical Surveys and Monographs, 139. American Mathematical Society, Providence, RI, 2007.

\bigskip

YURI I.~MANIN, 

{\it Max Planck Institute for Mathematics, Bonn}

manin\@mpim-bonn.mpg.de

\enddocument